\documentclass[12pt,a4paper]{article}
\usepackage{graphicx}
\usepackage{epsfig,cite}
\usepackage{epsfig,latexsym,fancyheadings,amssymb,feynmp}
\usepackage{amsmath}
\usepackage[english]{}

\catcode`@=11
\def\citer{\@ifnextchar
[{\@tempswatrue\@citexr}{\@tempswafalse\@citexr[]}}

\def\@citexr[#1]#2{\if@filesw\immediate\write\@auxout{\string\citation{#2}}\fi
  \def\@citea{}\@cite{\@for\@citeb:=#2\do
    {\@citea\def\@citea{--\penalty\@m}\@ifundefined
       {b@\@citeb}{{\bf ?}\@warning
       {Citation `\@citeb' on page \thepage \space undefined}}%
\hbox{\csname b@\@citeb\endcsname}}}{#1}} \catcode`@=12

\newcommand{\ML}{\left( \begin{array}{cc}}
\newcommand{\MR}{\end{array} \right)}

\newcommand{\BE}{\begin{equation}}
\newcommand{\EE}{\end{equation}}
\newcommand{\BEA}{\begin{eqnarray}}
\newcommand{\EEA}{\end{eqnarray}}

\newcommand{\tg}{\mathrm{tg}}
\newcommand{\ctg}{\mathrm{ctg}}

\begin{document}

\thispagestyle{empty}
\setcounter{page}{0}
\def\thefootnote{\fnsymbol{footnote}}

\, \mbox{}\vspace{30mm}
\begin{center}

{\large\sc {\bf Yukawa radiative corrections to the trilinear
self-couplings of}} \\[2mm]
{\large\sc {\bf  neutral \textbf{\textrm{CP}}-even Higgs bosons
and decay width $\Gamma(H \to hh)$ }} \\[2mm]
{\large \sc {\bf in the MSSM}}

\vspace*{0.3cm}

\vspace{1.0cm}

{\sc Yu.P.~Philippov \footnote{ email:yuphil@ssu.samara.ru}}

\vspace*{0.8cm} Samara State University, 443011 Samara, Russia
\vspace*{1.0cm}
\end{center}

\begin{abstract}
The four self-couplings $\lambda_{hhh}$, $\lambda_{hhH}$,
$\lambda_{hHH}$, $\lambda_{HHH}$ and decay width $\Gamma(H \to
hh)$ are calculated taken into account one-loop
$t,b,c$\,-\,(s)quarks, $\tau$\,-\,(s)lepton corrections in the
framework of the Minimal Supersymmetric Standard Model (MSSM). By
the example of self-couplings dependencies from  $ \tg  \beta$ and
\textrm{$M_A$}, it is shown, that calculated corrections can give
the essential contribution to self-couplings and decay width
determined at one-loop level. The radiative corrections account
becomes necessary condition for per\-for\-mance of Higgs potential
reconstruction procedure and experimental confirmation of Higgs
mechanism.
\end{abstract}

\vspace*{5.0cm}

\def\thefootnote{\arabic{footnote}}
\setcounter{footnote}{0}

\newpage


\section{Introduction}

\hspace{5mm} Higgs Mechanism is a crucial element in construction
of modern realistic gauge models in quantum field theory
\citer{Goldstone1,Kibble1}. Today, however, the given mechanism
has not received direct experimental confirmation.  One of steps
in program of experimental confirmation of Higgs mechanism
realization in nature is an experimental determination of Higgs
self-couplings, predicted in framework of the model
\cite{Zerwas01}. The given task becomes more actual in
supersymmetric extensions of Standard model (SM) where the
structure of self-couplings is also defined  by supersymmetry
principles. The last ones also have no experimental confirmation.

There are many processes in the MSSM, which amplitudes and total
cross sections are defined by Higgs self-couplings. In the ideal
case all the processes could be realized on future high luminosity
colliders and it would be possible to determine self-couplings by
their measured cross sections. In practice, not all the cross
sections will be large enough, to be accessible experimentally. In
works \citer{djouadi2,osland} it has been shown, that Higgs
self-couplings $\lambda_{hhh}, \lambda_{hhH}$ have maximal areas
of sensitivity to determination in elementary processes at
high-energy $e^+ e^-$\,--\,colliders.

There is special interest to the trilinear self-coupling
$\lambda_{hhH}$, because one defines width of decay $H \rightarrow
hh$ (at tree level):
\begin{equation}\label{5eq:HDecayRate}
\Gamma(H \to hh)=\frac{\lambda_{hhH}^2}{32 \pi M_H}\sqrt{1 -
\frac{4 M_h^2}{M_H^2}},
\end{equation}
where $M_{\{h,H\}}$\,--\,masses of CP-even Higgs bosons $h,H$. The
given channel is the basic mode of heavy Higgs boson decay in part
of the parameter space with $H$ masses between 200 and 350~GeV and
for moderate values of $\tg \beta$ \cite{zerwas1}. The resonant
$H$ decay enhances the production cross sections for
$e^+e^-$\,-\,processes with $hh$-final states by an order of
magnitude \cite{djouadi2} thus improving the potential for the
measurement of the Higgs self-coupling $\lambda_{Hhh}$. Moreover
this decay increases the cross sections of LHC processes with the
same final states by up to 2 orders of magnitude \cite{djkalzer}.

Necessity for the precise theoretical prediction of possible
values of Higgs self-couplings has pushed many experts to
calculation the parameters taken into account one-loop corrections
in various perturbation approaches. For instance, Higgs
self-couplings have been calculated in Renormalization Group
Approach (RGA) \citer{Self-coupl-1,Self-coupl-1-4}, taken into
account leading logarithmic $t-\tilde {t}$ one-loop corrections;
in Effective Potential Approach (EPA) with using $t-\tilde{t}$
one-loop corrections \citer{Self-coupl-2} and $t-\tilde{t}$,
$b-\tilde{b}$ one-loop corrections \citer{Self-coupl-2-2}; in
Feynman Diagram Approach (FDA)  taken into account both
$t-\tilde{t}$ one-loop corrections \cite{Self-coupl-3} and
complete set of one-loop corrections for
$\lambda_{hhh},\lambda_{HHH}$ self-couplings
\citer{Philippov_10,Philippov_11}.  It is shown, that
$b-\tilde{b}$ one-loop corrections are essential at large $\tg
\beta$ and should be taken into account. Summarizing the previous
work, it is possible to approve, that one-loop radiative
corrections to the trilinear Higgs self-couplings with lightest
Higgs boson ($h$) can be rather significant and their account
essentially modifies final results for observables.

The decay width $\Gamma(H \rightarrow hh)$ has been calculated
with use of one-loop results for Higgs self-couplings \cite
{osland, Brign-Zwin}. It is demonstrated, that calculated
corrections to self-couplings essentially increase values of
$\Gamma (H \rightarrow hh)$ for $M_H \leq 500$~GeV and moderate
values of $\tg \beta$.

The main goal of the present work  is calculation of  four Higgs
self-couplings  $\lambda_{hhh}$, $\lambda_{hhH}$, $\lambda_{hHH}$,
$\lambda_{HHH}$ and decay width $\Gamma(H \to hh)$ taken into
account Yukawa one-loop $t,b,c$\,-\,(s)quarks,
$\tau$\,-\,(s)lepton corrections in FDA. We consider that account
of $\tau-\tilde{\tau}$-, $c-\tilde{c}$\,-\,loops is expedient,
because Yukawa couplings of $b$-quark and $\tau$-lepton have
identical structure and masses of these particles and $c$-quark
have one order. We neglect Yukawa contributions of other particles
since their masses much less than masses specified particles.
Interest to the goal is caused by the analysis of Higgs potential
modification  by means of the one-loop corrections account (not
considered earlier), and use \textrm{on-shell} renormalization
scheme, based on the results of work \cite{Dabelstein-95}. The
scheme has not been used in calculations of radiative corrections
for Higgs self-couplings in general case.

\section{MSSM Higgs sector at tree level}
\hspace{5mm} The Higgs sector in MSSM includes two doublets of
scalar fields \footnote{\hspace{-1mm}$^{)}$Hereinafter we use the
designations offered in \cite{HHG}.}\hspace{0mm}$^{)}$:
\[
\begin{array}{cc}
H_1\!= \! \begin{pmatrix} \! \frac{1}{\sqrt{2}} (v_1 + \phi_1^{0} - i\chi_1^{0})  \\
  -\phi_1^- \!  \end{pmatrix}, \hspace{3mm} H_2\!=\! \begin{pmatrix} \!  \phi_2^+
\\ \frac{1}{\sqrt{2}}(v_2 + \phi_2^0 + i\chi_2^0) \end{pmatrix}.
\end{array}
\label{eq:Hgfields}
\]
They are characterized by hypercharge $Y_1=-Y_2=-1$ and vacuum
expectation values $v_1$, $v_2$. A part of lagrangian (Higgs
potential) contains mass terms and terms of scalar fields
interactions is determined by next expression
\begin{eqnarray}
V_{\rm Higgs} &=& m_1^2\,|H_1|^2+m_2^2\,|H_2|^2 - m_{12}^2\,\left(
\epsilon_{ij}\,H_1^i\,H_2^j+{\rm h.c.}\nonumber\right) + \\
&+& \frac{1}{8} (g_1^2+g_2^2)\,\left(|H_1|^2-|H_2|^2\right)^2
+\frac{1}{2}\,g_2^2\,|H_1^{\dagger}\,H_2|^2, \label{eq:potential}
\end{eqnarray}
ãäå $m_1^2$, $m_2^2$, $m_{12}^2$\,--\,soft SUSY-breaking
parameters, $g_1$, $g_2$\,--\,$U_{Y}(1)$ and $SU_L(2)$\,--\,gauge
constants;
$\epsilon_{11}=\epsilon_{22}=0,\,\epsilon_{12}=-\epsilon_{21}=1$.

For observables calculation it is necessary to proceed to basis of
physical fields by means of the following rotations in initial
scalar fields space (Higgs potential diagonalization):
\begin{equation}
\begin{pmatrix}
  \phi_1^0 \\ \phi_2^0
\end{pmatrix}  = D(\alpha)
\begin{pmatrix} H^0 \\ h^0 \end{pmatrix},\hspace{1mm} \begin{pmatrix} \chi_1^0 \\ \chi_2^0 \end{pmatrix}
= D(\beta)
\begin{pmatrix} G^0\\A^0\end{pmatrix},
\hspace{1mm}
\begin{pmatrix} \phi_1^{\pm} \\ \phi_2^{\pm} \end{pmatrix} = D(\beta)
\begin{pmatrix}  G^{\pm} \\ H^{\pm} \end{pmatrix}.
\label{HiggsFTrans}
\end{equation}
In expression (\ref{HiggsFTrans}) $D(\alpha),
D(\beta)$\,--\,matrixes of $O(2)$\,--\,rotations, $G^0$,
$G^{\pm}$--\,Goldstone modes. The MSSM Higgs physical states are
represented by: 1) two $CP$-even neutral states $\{h^0,H^0\}$, 2)
one $CP$-odd neutral state $A^0$, and  3) two charge states $
H^\pm$.

Higgs potential (\ref{eq:potential}) contains two free parameters
after diagonalization (\ref{HiggsFTrans}): a tangent of mixing
angle $\beta$ and $A^0$-boson mass, $M_A$ which are defined by the
following expressions:
\[ \label{freepar} \tg \beta =
\frac{v_2}{v_1}, \hspace{5mm} M_{A}^2 = m_{12}^2(\tg \beta + \ctg
\beta). \]
 The masses of other Higgs states are represented as follows
\[
m_{h,H}^2\! =\! \frac{1}{2} \left[ M_{A}^2 + M_Z^2 \mp
 \sqrt{(M_{A}^2+M_Z^2)^2
-4M_Z^2M_{A}^2\cos^22\beta}\hspace{1mm}\right]\!,\hspace{2mm}
m_{H^\pm}^2 \!= \! M_{A}^2 + M_W^2,
\]
ãäå $M_{\{W,Z\}}$\,--\,masses of $W$-,$Z$-bosons. Mixing angles
$\alpha$ and $\beta$ are connected by next expression
\[\label{mixanglconnect}
 \tg 2\alpha= \tg 2\beta \frac{M_A^2+M_Z^2}{M_A^2-M_Z^2},
 \hspace{5mm} -\frac{\pi}{2} < \alpha < 0.
\]
A part of Higgs sector lagrangian, included triple Higgs bosons
($h^0,H^0$) interactions looks as follows
\begin{eqnarray} \mathcal{L}_{Int}^{(3)}\!\!  &=&\!\!
\frac{\lambda_{hhh}^{(0)}}{3!} h^0h^0h^0 +
\frac{\lambda_{hhH}^{(0)}}{2!} h^0h^0H^0 +
\frac{\lambda_{hHH}^{(0)}}{2!} h^0H^0H^0 +
\frac{\lambda_{HHH}^{(0)}}{3!} H^0H^0H^0. \label{eq:TripleInt}
\end{eqnarray}
The main parameters characterized intensities of Higgs boson
 interactions are Higgs self-couplings $\{\lambda_{ijk}^{(0)}\}$.
The trilinear Higgs self-couplings for given bosons at tree level
are:
\begin{eqnarray}
\left. \begin{array}{ll} \lambda_{hhh}^{(0)} =  3 \lambda_0
s_{(\beta+\alpha)} c_{2\alpha}, & \lambda_{hhH}^{(0)} = \lambda_0
[2
s_{2\alpha} s_{(\alpha+\beta)} - c_{2\alpha} c_{(\alpha+\beta)}], \\[2mm]
\lambda_{hHH}^{(0)} =  - \lambda_0 [2 c_{(\beta+\alpha)}
s_{2\alpha} + s_{(\beta+\alpha)} c_{2\alpha}], &
\lambda_{HHH}^{(0)}  =  3 \lambda_0 c_{(\beta+\alpha)}
c_{2\alpha}, \end{array} \right\}, \label{Systemcouplings}
\end{eqnarray}
where $\lambda_0 = M_Z^2/v$, $v=\sqrt{v_1^2+v_2^2}$, $s_x=\sin
x,\,c_x=\cos x$. The Higgs self-coupling structure is modified
when we include radiative corrections in view.


\section{Higgs self-couplings calculation at one-loop level}

\subsection{Vertex functions at one-loop level}
\label{sec:VF1loop}
\hspace{5mm} The central object of further investigations is
$n$-point Vertex Function (VF), which in one-loop approximation is
represented as
\begin{equation}
\Gamma^{(n)}_{[\mathrm{R}]}= \Gamma^{(n)}_{[\mathrm{0}]} +
\hat{\Gamma}^{(n)}_{[\mathrm{1}]}, \label{eq:VFN}
\end{equation}
where first term $\Gamma^{(n)}_{[\mathrm{0}]}$ determines VF at
tree level, (for $n=3$ it coincides with self-coupling of system
(\ref{Systemcouplings})). Second term in (\ref{eq:VFN}) defines
renormalized one-loop contribution in vertex function and is shown
as sum of all considered one-loop corrections (summation on types
of diagrams ($i$) and on sets of virtual particles
fields\,--\,$\{j\}$) and counterterm:
\[ \label{1loopcor}
\hat{\Gamma}^{(n)}_{[\mathrm{1}]} = \sum_{i,\{j\}}\Delta
\Gamma_{i,\{j\}}^{(\mathrm{n})}+ \delta
 \Gamma^{(n)}.
\]

\begin{figure}[!h]
\vspace{-5mm}
\begin{center}
\includegraphics[width=0.95\textwidth]{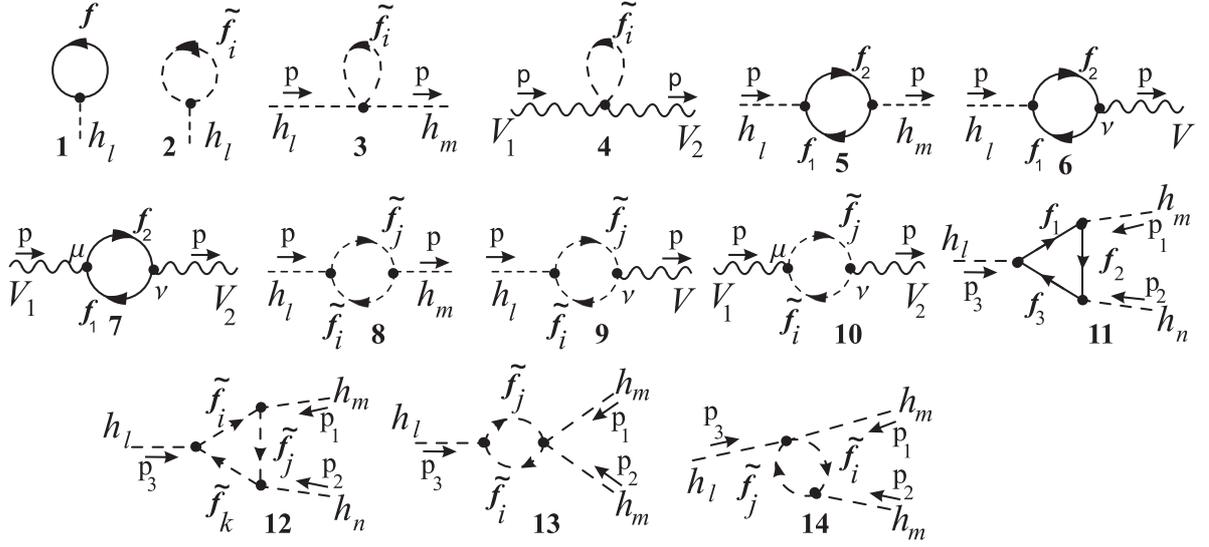}
\end{center}
\vspace{-7mm} \caption{generic Yukawa one-loop diagrams
contributing to the one-, two- and three-point Vertex Functions.}
\label{Fig:Setdiag}
\end{figure}

All types of one-loop diagrams contributing to the one-, two- and
three-point Vertex Functions are submitted on fig.
\ref{Fig:Setdiag}. Calculation of corrections is fulfilled in t'
Hooft-Feynman gauge with use of a standard set of Feynman rules
\cite{HHG}.

\subsection{\textrm{On-shell}\,-\,renormalization of  vertex function}
\hspace{5mm} For definition of counterterm $\delta
 \Gamma^{(n)}$ we shall use standard
\textrm{On-shell} renormalization scheme \cite{Dabelstein-95}.
Structure of counterterms is fixed by system of standard
renormalization conditions for
\begin{itemize}
    \item self-energies of gauge bosons $Z,\,W, \gamma$, and
Higgs boson $A$;
    \item mixing energies $A-Z$,$\gamma-Z$;
    \item  residue conditions for $A,\,\gamma$-bosons propagators;
    \item the renormalization of $\tg \beta$ in such a way that the
    relation  $\tg \beta = v_2/v_1$ is valid for the one-loop Higgs minima;
    \item the tadpole conditions for vanishing renormalized tadpoles, i.e.
the sum of the one-loop tadpole diagrams for $h^0$, $H^0$ and the
corresponding tadpole counterterm is equal to zero.
\end{itemize}
Solving the received system of 11 linearized equations concerning
initial counterterms, we have received the following results for
counterterms of three-point Higgs Vertex functions:
\begin{eqnarray}
&&\!\!\!\! \textstyle \delta \Gamma_{hhh}= \frac{3 g_2 M_Z^2}{4
M_W^2} \left[\Big(-3 M_W \Sigma_A'(M_A^2) + \frac{1}{M_W
s_w^2}(1-2 s_w^2)\Sigma_W(M_W^2) + \frac{M_W}{M_Z^2 s_w^2}(-1 + 3
s_w^2)\times  \right. \nonumber \\[1mm]
&&\!\!\!\!  \textstyle  \times \Sigma_Z(M_Z^2) + M_W
\Sigma_{\gamma}'(0) - \frac{2 s_w}{M_Z} \Sigma_{\gamma Z}(0)\Big)
c_{2 \alpha} s_{(\alpha + \beta)} + \frac{M_W}{2 M_Z s_{2 \beta}}
\Sigma_{AZ}(M_A^2) \Big(- s_{( \alpha - 3\beta)} +   \nonumber \\[1mm]
&&\!\!\!\! \left. \textstyle + 3 s_{(3 \alpha - \beta)}  - 7
s_{(\alpha
+ \beta)} + s_{(3(\alpha + \beta))}\Big) \right], \nonumber \\[2mm]
&&\!\!\!\! \textstyle \delta \Gamma_{hhH} = \frac{g_2M_Z^2}{4
M_W^2} \left[ \Big(- \frac{3}{2} M_W \Sigma_A'(M_A^2) - \frac{1}{2
M_W s_w^2}(-1 + 2s_w^2)\Sigma_W(M_W^2) + \frac{M_W}{2 M_Z^2
s_w^2}(-1 + 3s_w^2) \times
 \right. \nonumber \\[1mm]
&&\!\!\!\! \textstyle \times \Sigma_Z(M_Z^2)  + \frac{M_W}{2}
\Sigma_{\gamma}'(0) - \frac{s_w}{M_Z} \Sigma_{\gamma Z}(0)\Big)
(c_{(\alpha - \beta)} - 3 c_{(3 \alpha + \beta)}) + \frac{M_W}{2
M_Z s_{2 \beta}} \Sigma_{AZ}(M_A^2) \Big( c_{(\alpha - 3\beta)} -
  \nonumber \\[1mm]
&&\!\!\!\! \textstyle \left. - 9 c_{(3 \alpha - \beta)} +7
c_{(\alpha + \beta)} - 3 c_{(3(\alpha + \beta))} \Big) \right],
\nonumber
\end{eqnarray}
\begin{eqnarray}
&&\!\!\!\! \textstyle \delta \Gamma_{hHH} = \frac{g_2 M_Z^2}{4
M_W^2} \left[ \Big(\frac{3}{2} M_W \Sigma_A'(M_A^2) + \frac{1}{2
M_W s_w^2}(-1 + 2s_w^2)\Sigma_W(M_W^2) + \frac{M_W}{2 M_Z^2
s_w^2}(-1 + 3s_w^2) \times \right. \nonumber \\[1mm]
&&\!\!\!\! \textstyle \times \Sigma_Z(M_Z^2) + \frac{M_W}{2}
\Sigma_{\gamma}'(0) + \frac{s_w}{M_Z}\Sigma_{\gamma Z}(0) \Big)
(s_{(\alpha - \beta)} + 3 s_{(3 \alpha + \beta)}) - \frac{M_W}{2
M_Z s_{2 \beta}} \Sigma_{AZ}(M_A^2) \Big(s_{(\alpha - 3\beta)} +
\nonumber  \\[1mm] &&\!\!\!\! \textstyle \left. + 9 s_{(3 \alpha - \beta)} +
7 s_{(\alpha + \beta)} + 3 s_{(3(\alpha + \beta))}\Big) \right], \nonumber \\[1mm]
&&\!\!\!\! \textstyle \delta \Gamma_{HHH}= \frac{3 g_2 M_Z^2}{4
M_W^2} \left[\Big(-3 M_W \Sigma_A'(M_A^2) + \frac{1}{M_W
s_w^2}(1-2 s_w^2)\Sigma_W(M_W^2) + \frac{M_W}{M_Z^2 s_w^2}(-1 + 3
s_w^2) \times \right. \nonumber \\[1mm]
&&\!\!\!\! \textstyle \times \Sigma_Z(M_Z^2) + M_W
\Sigma_{\gamma}'(0) - \frac{2 s_w}{M_Z} \Sigma_{\gamma Z}(0)\Big)
c_{2 \alpha} c_{(\alpha + \beta)} + \frac{M_W}{2 M_Z s_{2 \beta}}
\Sigma_{AZ}(M_A^2)\Big(c_{(\alpha - 3\beta)} +  \nonumber \\[1mm]
&&\!\!\!\! \left.  \textstyle + 3 c_{(3 \alpha - \beta)} + 7
c_{(\alpha + \beta)} + c_{(3(\alpha + \beta))}\Big) \right].
\nonumber
\end{eqnarray}
In last system $s_w =\sin \theta_w$, $\theta_w$\,--\,Weinberg
angle.
\subsection{Decay Amplitude calculation at one-loop level}
\hspace{5mm} \label{sec:OneLoopAmp} Calculation of decay width at
one-loop level is reduced to calculation of process amplitude
$\mathcal{A}_{H \to hh}$, since one defines expression for decay
width as
\begin{equation}\label{DecayRate}
\Gamma(H \rightarrow hh) = \frac{\left|\mathcal{A}_{H \to hh}
\right|^2}{32 \pi M_H} \sqrt{1 - \frac{4 M_h^2}{M_H^2}},
\end{equation}
where $M_h, M_H$\,--\,masses of $CP$-even Higgs physical states
Õèããñà in the corresponding approximation, being real parts of the
following equation roots:
\[
 \left[ p^2 - m^2_{h} + \hat{\Sigma}_{hh} (p^2) \right]
\left[ p^2 - m^2_{H} + \hat{\Sigma}_{HH} (p^2) \right] - \left[
\hat{\Sigma}_{hH}
 (p^2)\right]^2 = 0.
\label{PolesSolution}
\]
For determination of $\mathcal{A}_{H \to hh}$ we shall make
transition to basis of renormalized fields by means of $Z$-matrix
\[
\begin{pmatrix}
  h \\
  H \\
\end{pmatrix}
\mapsto
\begin{pmatrix}
  Z_{hh}^{\frac{1}{2}} & Z_{hH}^{\frac{1}{2}} \\
  Z_{Hh}^{\frac{1}{2}} & Z_{HH}^{\frac{1}{2}} \\
\end{pmatrix}
\begin{pmatrix}
  h_R \\
  H_R \\
\end{pmatrix}
\]
in lagrangian (\ref {eq:TripleInt}) in which are included Higgs
self-couplings at one-loop level. Isolating the coefficient at
product $h_Rh_RH_R $, we derive expression for amplitude of
process
\begin{equation}\label{Amplitude1}
{\cal A} = Z^{\frac{1}{2}}_{HH}\left[Z^{\frac{1}{2}}_{hh}\right]^2
\Gamma_{hhH}^{[I]} + 2 Z^{\frac{1}{2}}_{HH} Z^{\frac{1}{2}}_{hh}
Z^{\frac{1}{2}}_{Hh} \Gamma_{hHH}^{[I]} +
Z^{\frac{1}{2}}_{hH}\left[Z^{\frac{1}{2}}_{hh}\right]^2
\Gamma_{hhh}^{[I]}.
\end{equation}
Taking into account next on-shell condition: real part of residue
from propagator in a pole  should be equal to unit, it is possible
to show, that $Z$-matrix elements are represented as follows:
\begin{displaymath}\label{Resolution1}
\begin{array}{ll}
Z^{\frac{1}{2}}_{hh}= 1-\hat{\Sigma}\,'_{hh}(M_h^2),\hspace{3mm} &
Z^{\frac{1}{2}}_{hH}= -\hat{\Sigma}_{hH}(M_H^2)/(M_H^2-M_h^2),\\[2mm]
Z^{\frac{1}{2}}_{Hh}=
-\hat{\Sigma}_{hH}(M_h^2)/(M_h^2-M_H^2),\hspace{3mm} &
Z^{\frac{1}{2}}_{HH}=  1-\hat{\Sigma}\,'_{HH}(M_H^2).
\end{array}
\end{displaymath}

\section{Numerical results and analysis}

\hspace{5mm} Let's consider numerical results for self-couplings
and decay width derived by scheme of calculation, presented in
previous sections.
\begin{figure}[!b]
\begin{center}
\includegraphics[width=0.95\textwidth]{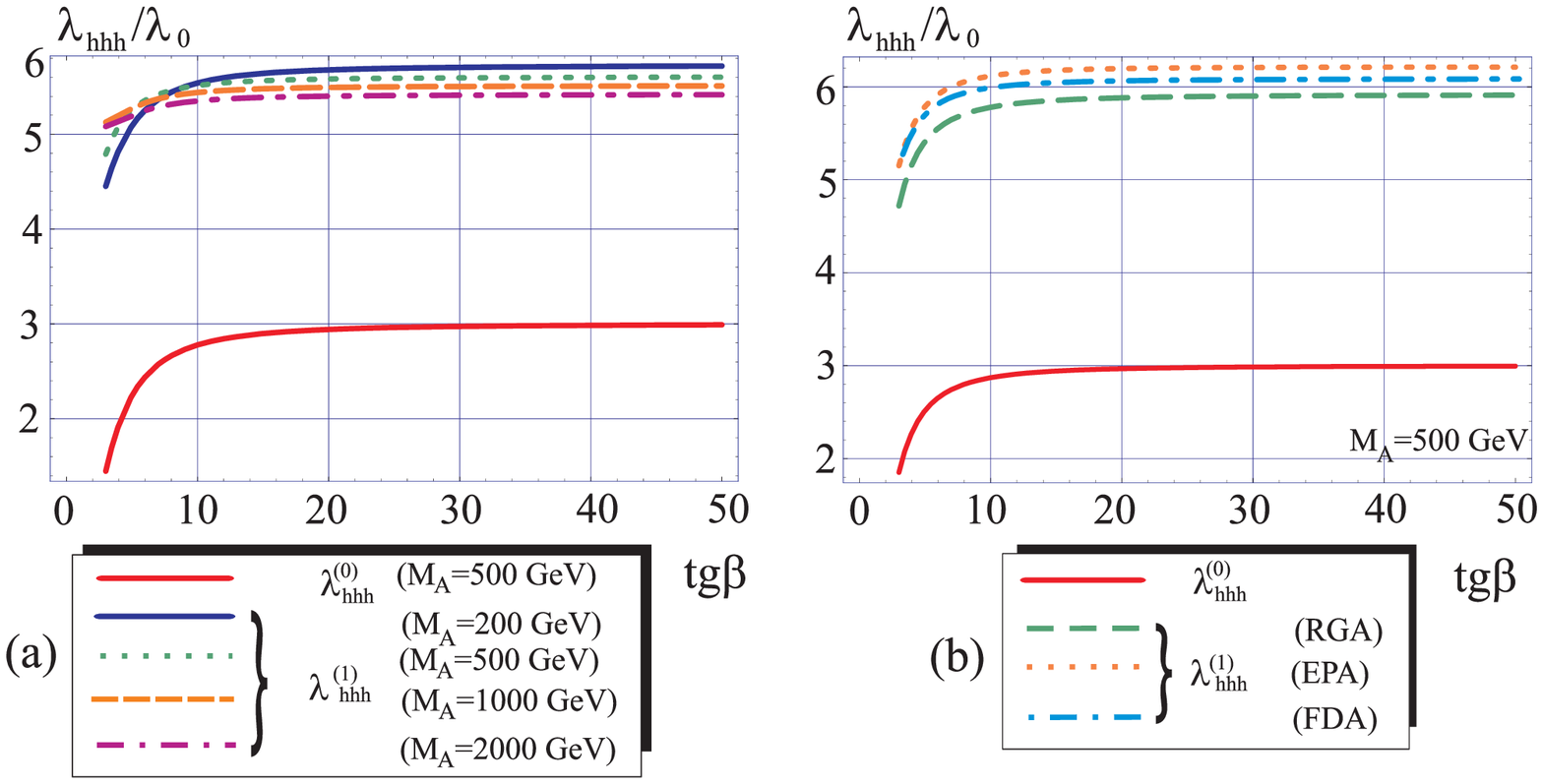}
\end{center}
\vspace{-6mm} \caption{dependence from $\tg \beta$ of  (a) new
results for self-coupling $\lambda_{hhh}$ (in units $
\lambda_{0}$); (b) RGA results
\cite{Self-coupl-1-2,Self-coupl-1-3,Self-coupl-1-4}, EPA results
\cite{Self-coupl-2}, FDA complete one-loop results
\cite{Philippov_10,Philippov_11} for $A_{f}=1$~TeV, $\mu=0.5$ TeV,
$M_{\tilde{Q}} =$ $M_{\tilde{U}} =$ $M_{\tilde{D}} =$
$M_{\tilde{L}} =$ $M_{\tilde{R}}= 1$ ~TeV, $\sqrt{p_3^2} = (2 M_h
+ 10)$~GeV.} \label{Fig:lhhh}
\end{figure}
\begin{figure}[!b]
\begin{center}
\includegraphics[width=0.95\textwidth]{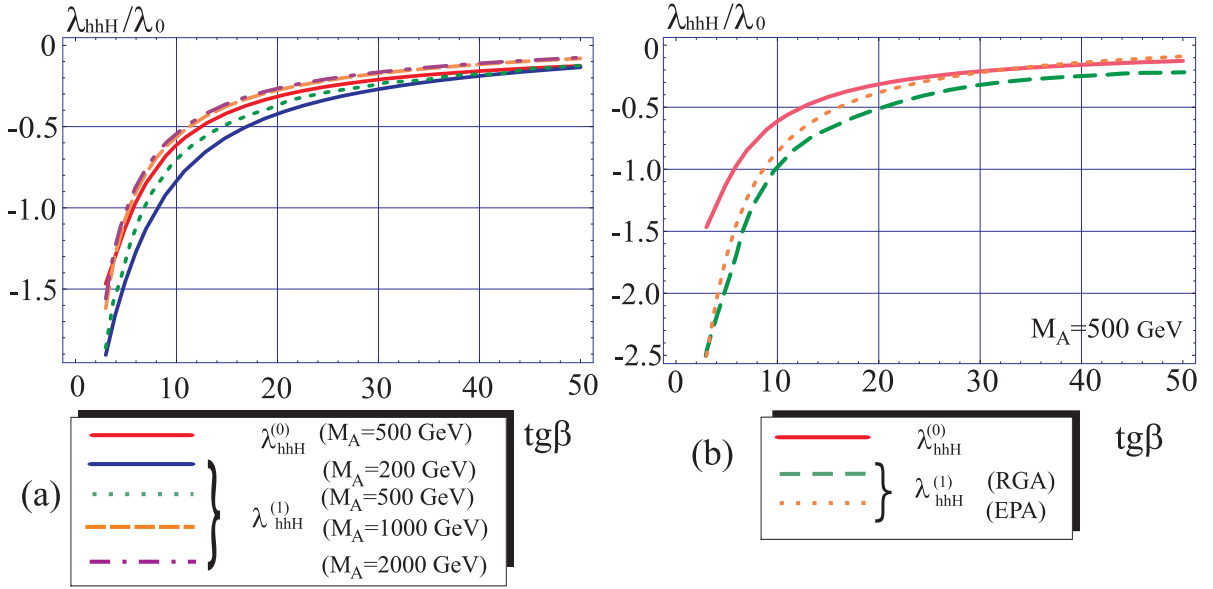}
\end{center}\vspace{-6mm}
\caption{dependence from $\tg \beta$ of  (a) new results for
self-coupling $\lambda_{hhH}$ (in units $ \lambda_{0}$); (b) RGA
results \cite{Self-coupl-1-2,Self-coupl-1-3,Self-coupl-1-4}, EPA
results \cite{Self-coupl-2} for $A_{f}=1$~TeV, $\mu=0.5$ TeV,
$M_{\tilde{Q}} =$ $M_{\tilde{U}} =$ $M_{\tilde{D}} =$
$M_{\tilde{L}} =$ $M_{\tilde{R}}= 1$ ~TeV, $\sqrt{p_3^2} =(2M_h +
10)$~GeV.} \label{Fig:lhhH}
\end{figure}
Dependence  $\lambda_{hhh}$ (in units $ \lambda_{0}= M_Z^2/v$)
from $\tg \beta$ for four values of $M_A$ is represented on fig.
\ref{Fig:lhhh}.(a). Obviously, one-loop contribution essentially
modifies tree-level result for any value $M_A$. The problem of
significant corrections to the given self-coupling has been
investigated in details in \cite{Self-coupl-3}; the exhaustive
explanation to the specified phenomenon has been given. On fig.
\ref{Fig:lhhh}.(b) RGA results
\cite{Self-coupl-1-2,Self-coupl-1-3,Self-coupl-1-4}, EPA results
\cite{Self-coupl-2}, FDA complete one-loop results
\cite{Philippov_10,Philippov_11} are shown. The new results are
most consistent with RGA results. Difference between new results
and FDA complete one-loop results \cite{Philippov_10,Philippov_11}
is determined by non-leading one-loop corrections with virtual
gauge bosons, Higgs bosons, chargino, neutralino (discrepancy is
about $\leq 0.25 \lambda_0$). All investigating Higgs
self-couplings  very weakly depend from $M_A, \mu, A_f, M_S$,
therefore we will not demonstrate  these dependencies.
\begin{figure}[!h]
\begin{center}
\includegraphics[width=0.95\textwidth]{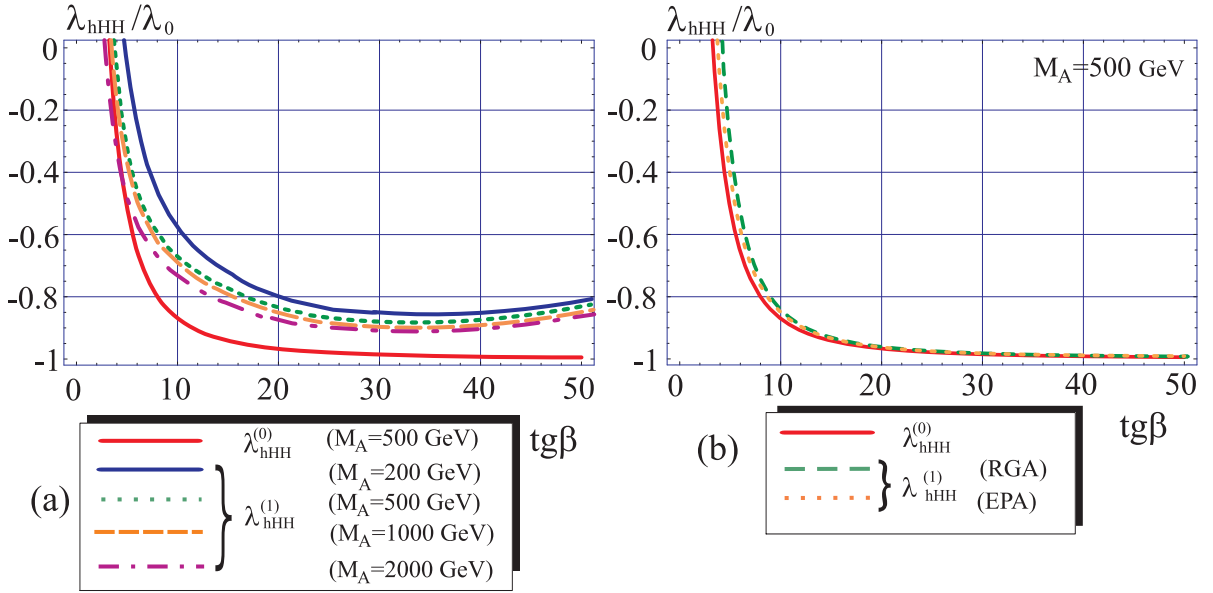}
\end{center}\vspace{-6mm}
\caption{dependence from $\tg \beta$ of  (a) new results for
self-coupling $\lambda_{hHH}$ (in units $ \lambda_{0}$); (b) RGA
results \cite{Self-coupl-1-2,Self-coupl-1-3,Self-coupl-1-4}, EPA
results \cite{Self-coupl-2} for $A_{f}=1$~TeV, $\mu=0.5$ TeV,
$M_{\tilde{Q}} =$ $M_{\tilde{U}} =$ $M_{\tilde{D}} =$
$M_{\tilde{L}} =$ $M_{\tilde{R}}= 1$ ~TeV, $\sqrt{p_3^2} =(M_h +
M_H + 10)$~GeV.} \label{Fig:lhHH}
\end{figure}
\begin{figure}[!h]
\begin{center}
\includegraphics[width=0.95\textwidth]{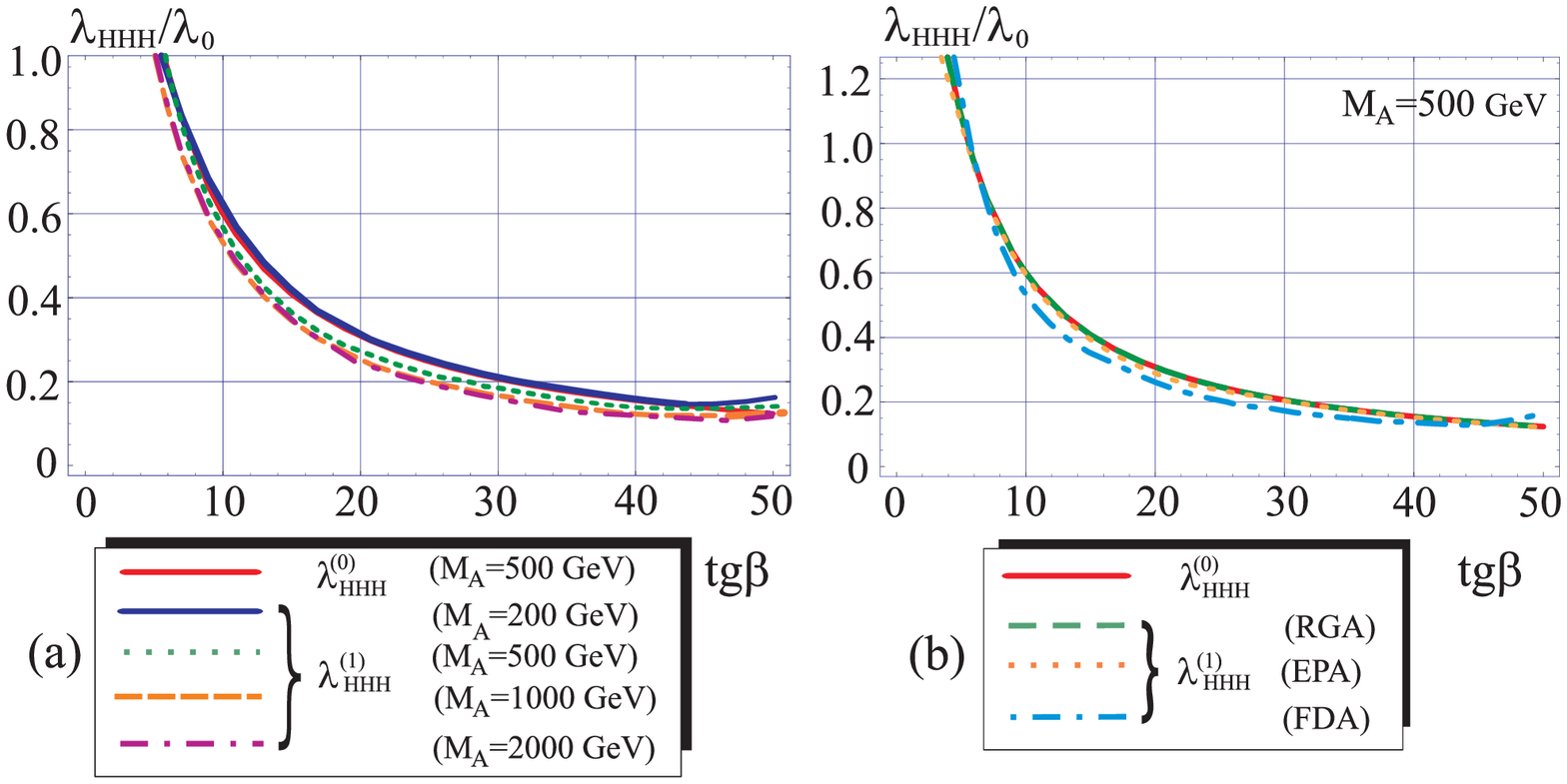}
\end{center}
\vspace{-6mm} \caption{dependence from $\tg \beta$ of  (a) new
results for self-coupling $\lambda_{HHH}$ (in units $
\lambda_{0}$); (b) RGA results
\cite{Self-coupl-1-2,Self-coupl-1-3,Self-coupl-1-4}, EPA results
\cite{Self-coupl-2}, FDA complete one-loop results
\cite{Philippov_10,Philippov_11} for $A_{f}=1$~TeV, $\mu=0.5$ TeV,
$M_{\tilde{Q}} =$ $M_{\tilde{U}} =$ $M_{\tilde{D}} =$
$M_{\tilde{L}} =$ $M_{\tilde{R}}= 1$ ~TeV, $\sqrt{p_3^2} =(2M_H +
10)$~GeV.} \label{Fig:lHHH}
\end{figure}
\begin{figure}[!h]
\begin{center}
\includegraphics[width=0.95\textwidth]{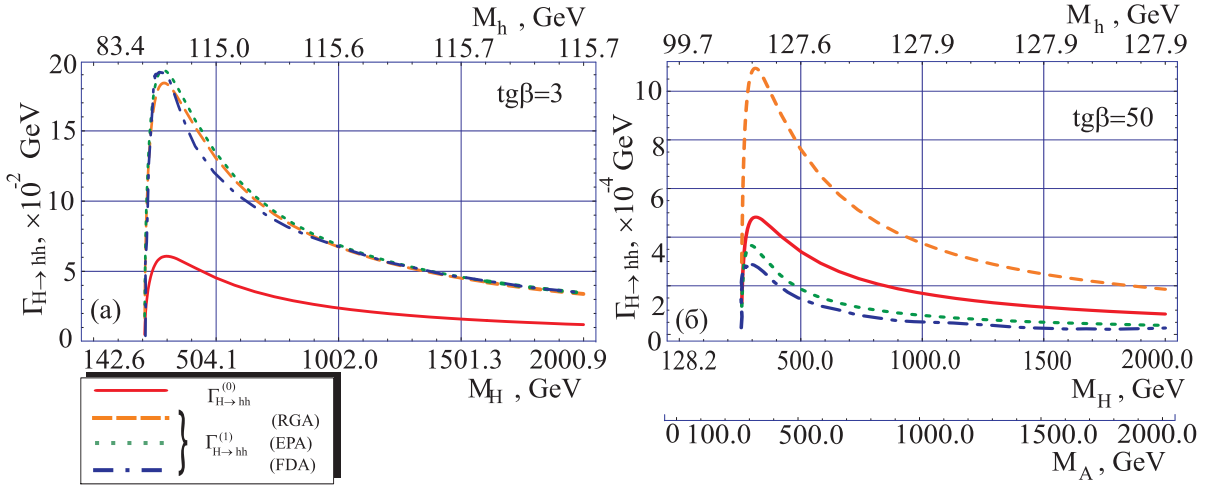}
\end{center} \vspace{-6mm}
\caption{dependence of decay width from Higgs boson mass
($M_h$,$M_H$, $M_A$). for (a) $\tg \beta = 3$, (b)  $\tg \beta =
50$ and  $A_{f}=1$~TeV, $\mu=0.5$ TeV, $M_{\tilde{Q}} =$
$M_{\tilde{U}} =$ $M_{\tilde{D}} =$ $M_{\tilde{L}} =$
$M_{\tilde{R}}= 1$~TeV.} \label{Fig:DRHtohh}
\end{figure}

On fig. \ref{Fig:lhhH}(a)-(b) the curves of dependence
$\lambda_{hhH}(\tg \beta)$ are demonstrated. Apparently, one-loop
contributions in this case are not so large as in previous case.
Maximal value of correction is reached for small values $M_A$ and
$\tg \beta$. It is necessary to note, that our results in the
specified area of values $M_A $ and $\tg \beta$ are less then
values RGA and EPA results. This fact is caused by nonzero value
$p_3$, impulse of virtual Higgs boson.

Fig. \ref{Fig:lhHH} demonstrates the curves of dependence
$\lambda_{hHH}(\tg \beta)$, which are given in different
perturbative approaches. Our curves testify to significant size of
the corrections as against predecessors results. The reason for
that -- display of threshold effects. It is well known, correct
description of observables near threshold of production is
achieved only in Feynman diagram approach. Our results are derived
at $\sqrt{p_3^2}=(M_h + M_H + 10)$~GeV, that corresponds to
threshold region.  There is one more feature -- at large $\tg
\beta$ we can observe significant increase of Higgs self-coupling,
calculated at one-loop level. This is a result of
$b\tilde{b}$,\,$\tau\tilde{\tau}$\,--\,loop corrections growth.
The last fact is bright confirmation of necessity in the given
one-loop contributions account.

As for one-loop corrections for self-couplings $\lambda_{HHH}$,
that ones are small. Growth of loop corrections at large $\tg
\beta$ and threshold effect are not shown almost, since
calculation is carried out for more heavy particles.

The dependence of decay width $\Gamma(H \to hh)$ from mass of
neutral Higgs boson(s) at various values $\tg \beta$ is submitted
on fig. \ref{Fig:DRHtohh}.(a)-(b). Fine agreement of different
approaches results  we can see at $\tg\beta=3$. The reason is
obvious, dominating contributions into final result for decay
width are $t\tilde{t}$\,--\,corrections which are taken into
account in all approaches. At $\tg \beta =50$ results EPA and FDA
are close.  Discrepancy is caused only by the account of $b\tilde
{b}$, $\tau\tilde{\tau}$\,-\,loop contributions in FDA.

\begin{figure}[!h]
\begin{center}
\includegraphics[width=0.95\textwidth]{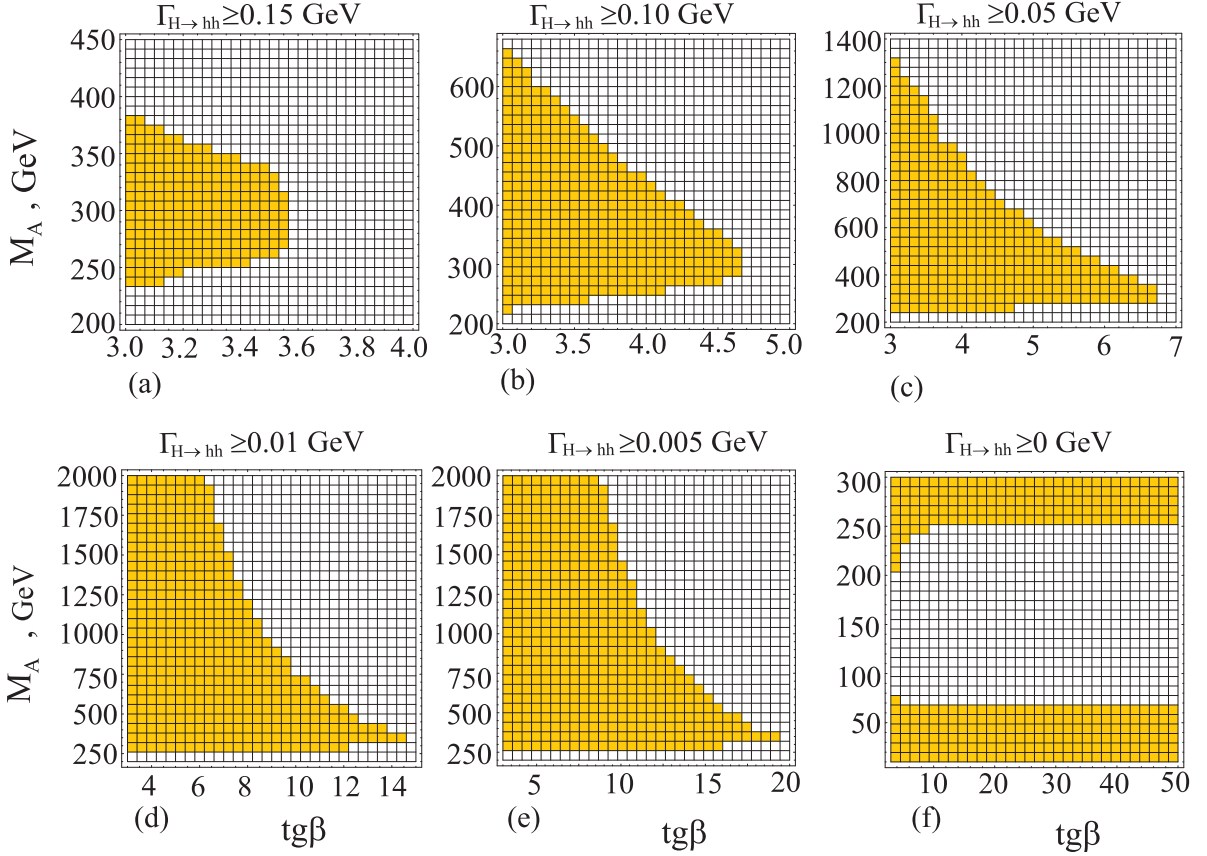}
\end{center}\vspace{-6mm}
\caption{(a)-(f) sensitivity areas of $\Gamma_{H \to hh}$ for fix
limit value, represented in plane $\tg \beta - M_A$, for
$A_{f}=1$~TeV, $\mu = 0.5$~TeV, $ M_{\tilde{Q}} = M_{\tilde{U}} =
M_{\tilde{D}} = M_{\tilde{L}} = M_{\tilde{R}}= 1$ ~TeV.}
\label{Fig:DRSAHtohh}
\end{figure}

Fig. \ref{Fig:DRSAHtohh}.(a)-(f) demonstrates sensitivity areas of
$\Gamma_{H \to hh}$ for fix limit value, represented in plane $\tg
\beta - M_A$.
It is necessary to note, that our results have "more good"
behaviour than outcomes of \cite{Brign-Zwin}, since the region
where $\Gamma(H \to hh) \geq 0.15$ (GeV) (in our results) lays in
the interval of heavier $M_A$. It is more preferable case for
recent experimental restrictions and theoretical scenarios.

\section{Conclusions}

\hspace{5mm} Thus dependencies of Higgs-self couplings
$\lambda_{hhh}$, $\lambda_{hhH}$, $\lambda_{hHH}$, $\lambda_{HHH}$
and decay width $\Gamma(H \to hh)$ from $\tg \beta$, $M_A$ have
been analysed in present work. It has been shown, that one-loop
results can essentially differ from tree-level ones. Applying for
the precision comparative analysis of theoretical and experimental
results with goal of self-couplings determination, it is necessary
to use at least one-loop approximation. Last fact becomes a
necessary condition of Higgs potential reconstruction and
experimental confirmation of Higgs mechanism realization in
nature.

\bibliographystyle{plain}

\begin{thebibliography}{99}
\bibitem{Goldstone1} Goldstone J. Field theories with "superconductor" solutions~//
               Nuovo Cimento. 1961. V.\,19. P.~154-164.
\bibitem{Nambu-Jona} Nambu Y., Jona-Lasinio G. Dynamical model of elementary
               particles based on an analogy with superconductivity.
               I~// Phys. Rev. 1961. V.\,122. P.~345-358.
\bibitem{Gold-Sal-Wein} Goldstone J., Salam A., Weinberg S. Broken
               symmetries~// Phys. Rev. 1962.  V.\,127. P.~965-970.
\bibitem{Higgs1} Higgs P.W. Broken symmetries, massless particles and gauge
                   fields~// Phys. Lett. 1964. V.\,12. P.~132-133.
\bibitem{Higgs2} Higgs P.W. Broken symmetries and the masses of gauge
                 bosons~// Phys. Rev. Lett. 1964.  V.\,13. P.~508-509.
\bibitem{Higgs3} Higgs  P.W. Spontaneous symmetry breakdown without massless
                 bosons~// Phys. Rev. 1966. V.\,145. P.~1156-1163.
\bibitem{Englert1}  R. Brout, E. Englert  Broken symmetry and the mass of gauge
                   vector mesons~// Phys. Rev. Lett. 1964. V.\,13. P.~321-322.
\bibitem{Gural-Hag-Kib} Guralnik G.S., Hagen C.R., Kibble T.W.B. Global conservation
                   laws and massless particles~// Phys. Rev. Lett. 1964. V.\,13.
                   P.~585-587.
\bibitem{Kibble1} Kibble T.W.B. Symmetry breaking in nonabelian gauge
                  theories~// Phys. Rev. 1967. V.\,155. P.~1554-1561.
\bibitem{Zerwas01} Zerwas P.M. Physics with an e+ e- linear collider at high
luminosity~// Eprint:hep-ph/0003221. 26pp.
\bibitem{djouadi2}   Djouadi A., Haber H. E., Zerwas P. M. Multiple production
                     of MSSM neutral higgs bosons at high-energy e+ e- colliders~//
                     Phys. Lett. B  1996. V.\,375. P.~203-212.
\bibitem{ours} Djouadi~A., Kilian~W., M\"uhlleitner~M. and Zerwas~P.M. Testing
               Higgs selfcouplings at e+ e- linear colliders~// Eur. Phys. J.
               1999. C\,10. P.~27-43.
\bibitem{ours-2} Djouadi~A., Kilian~W., M\"uhlleitner~M. and Zerwas~P.M. Production
                 of neutral higgs boson pairs at LHC~// Eur. Phys. J. 1999. C\,10.
                 P.~45-49.
\bibitem{osland}    Osland P., Pandita P. N. Measuring the trilinear couplings
                    of MSSM neutral higgs bosons at high-energy e+ e-
                    colliders~// Phys. Rev. 1999. D V. 59.
                    P.~055013. 18pp; Measuring trilinear Higgs couplings in the MSSM~//
                    E-print: hep-ph/9902270. 12pp; Multiple Higgs production and measurement
                    of Higgs trilinear couplings in the MSSM~//
                    E-print: hep-ph/9911295. 12pp.
\bibitem{zerwas1}     Djouadi~A., Kalinowski~J. and Zerwas~P.M. Two and three-body
                      decay modes of SUSY Higgs particles.~// Z. Phys. 1996.
                      C\,70. P.~435-448.
\bibitem{djkalzer}     Djouadi~A., Kalinowski~J. and Zerwas~P.M. Exploring
                       the SUSY Higgs sector at e+ e- linear colliders:
                       A Synopsis~// Z.\ Phys.\ 1993. C\,57 P.~569-584.
\bibitem{Self-coupl-1}
Haber~H.E. and Hempfling~R. Can the mass of the lightest Higgs
boson of the minimal supersymmetric model be larger than $M_Z$?
~// Phys.\ Rev.\ Lett.\ 1991. V.\,66. P.~1815-1818.
\bibitem{Self-coupl-1-2}Haber~H.E., Hempfling~R. Nir~Y. The decay $h_0\to A_0A_0$
in the minimal supersymmetric model~// Phys. Rev. D 1992. V.\,46.
P.~3015-3024.
\bibitem{Self-coupl-1-3} Ellis~J., Ridolfi~ G. and Zwirner~F. Radiative corrections
 to the masses of supersymmetric Higgs bosons~// Phys. Lett. B 1991. V.\,257.
 P.~83-91.
\bibitem{Self-coupl-1-4} Okada~Y., Yamaguchi~M. and
Yanagida~T. Upper bound of the lightest Higgs boson mass in the
minimal supersymmetric standard model~// Prog.\ Theor.\ Phys.
1991. V.\,85. P.~1-6.
\bibitem{Self-coupl-2}
Barger~V., Berger~M.\,S., Stange~A.\,L. and Phillips~R.\,J.\,N.
Supersymmetric higgs boson hadroproduction and decays including
radiative corrections~//Phys. Rev. D 1992. V.\,45. P.~4128-4147;
\bibitem{Self-coupl-2-2}
Kunszt~Z. and Zwirner~F. Testing the Higgs sector of the minimal
supersymmetric standard model at large hadron colliders~// Nucl.
Phys. B 1992. V.\,385. P.~3-75.
\bibitem{Self-coupl-3}
Hollik~W., Penaranda~S. Yukawa coupling quantum corrections to the
selfcouplings of the lightest MSSM Higgs boson~//Eur. Phys. J.
2002. C\,23. P.~163-172.
\bibitem{Philippov_10}
Dolgopolov M.V., Philippov~Yu.P. The trilinear neutral Higgs
self-couplings in the MSSM. Complete one-loop analysis~// Vestnik
of Samara State University~/ Samara, "Samara University", 2-nd
special realise 2003. P.87-95; hep-ph/0310263, 2003. 6pp.
\bibitem{Philippov_11}
Dolgopolov M.V., Philippov~Yu.P. Vertex functions of the
three-partial interaction of Higgs bosons $h^0$, $H^0$ in MSSM:
one-loop analysis~// Yad. Fiz. 2004. V.67. \symbol{157} 3.
P.~609-613.
\bibitem{Brign-Zwin} Brignole~A., Zwirner~F.  Radiative corrections to the
decay $H \to h h$ in the minimal supersymmetric standard model~//
Phys. Lett. B  1993. V.\,299.
 P.~72-82.
\bibitem{Dabelstein-95} Dabelstein A. The one loop renormalization of
the MSSM higgs sector and its application to the neutral scalar
higgs masses~// Z. Phys. 1995. C\,67. P.~495-512.
\bibitem{HHG} J.~F.~Gunion, H.~E.~Haber, G.~Kane, and  S.~Dawson,
The Higgs Hunter's Guide (Addison-Wesley, 1990).
\bibitem{Brignole92-1} Brignole~A. Radiative corrections to the supersymmetric
charged Higgs boson mass~// Phys. Lett. B 1992. V.\,277.
P.~313-323.
\bibitem{Brignole92-2} Brignole~A. Radiative corrections to the supersymmetric
neutral Higgs boson masses~// Phys. Lett. B 1992. V.\,281.
P.~284-294.
\end{thebibliography}



\end{document}